An insulating doped antiferromagnet with low magnetic symmetry as a room temperature spin conduit

Andrew Ross[1,2], Romain Lebrun[1,3], Lorenzo Baldrati[1], Akashdeep Kamra[4], Olena Gomonay[1], Shilei Ding[1,2,5], Felix Schreiber[1], Dirk Backes[6], Francesco Maccherozzi[6], Daniel A. Grave[7,8], Avner Rothschild[8], Jairo Sinova[1,9], Mathias Kläui[1,2,4*]

[1] *Institute of Physics, Johannes Gutenberg-University Mainz, 55099, Mainz, Germany*

[2] *Graduate School of Excellence Materials Science in Mainz, Staudingerweg 7, 55128, Mainz, Germany*

[3] *Unité Mixte de Physique CNRS, Thales, University Paris-Sud, Université Paris-Saclay, Palaiseau 91767, France*

[4] *Center for Quantum Spintronics, Department of Physics, Norwegian University of Science and Technology, NO-7491 Trondheim, Norway.*

[5] *State Key Laboratory for Mesoscopic Physics, School of Physics, Peking University, Beijing 100871, China*

[6] *Diamond Light Source, Didcot, Oxfordshire OX11 0DE, United Kingdom*

[7] *Department of Materials Engineering and Ilse Katz Institute for Nanoscale Science and Technology, Ben-Gurion University of the Negev, Beer-Sheva 8410501, Israel*

[8] *Department of Materials Science and Engineering, Technion-Israel Institute of Technology, Haifa 32000, Israel*

[9] *Institute of Physics ASCR, v.v.i., Cukrovarnicka 10, 162 53 Praha 6 Czech Republic*

*klaeui@uni-mainz.de

**Abstract**:

**We report room temperature long-distance spin transport of magnons in antiferromagnetic thin film hematite doped with Zn. The additional dopants significantly alter the magnetic anisotropies, resulting in a complex equilibrium spin structure that is capable of efficiently transporting spin angular momentum at room temperature without the need for a well-defined, pure easy-axis or easy-plane anisotropy. We find intrinsic magnon spin-diffusion lengths of up to 1.5 μm, and magnetic domain governed decay lengths of 175 nm for the low frequency magnons, through electrical transport measurements demonstrating that the introduction of non-magnetic dopants does not strongly reduce the transport length scale showing that the magnetic damping of hematite is not significantly increased. We observe a complex field dependence of the non-local signal independent of the magnetic state visible in the local magnetoresistance and direct magnetic imaging of the antiferromagnetic domain structure. We explain our results in terms of a varying and applied-field-dependent ellipticity of the magnon modes reaching the detector electrode allowing us to tune the spin transport.**

Antiferromagnetic (AFM) spintronics seeks to utilize the high-frequency dynamics, stability against magnetic perturbations and negligible stray fields in functionalizing AFM materials[1]. The electrical reading[2,3] and writing[4,5] of the Néel vector ***n*** in insulating AFMs, which further benefit from reduced Joule heating, demonstrate the role AFMs can play in developing devices, for instance, for memories. The Néel vector has been shown to efficiently transport AFM magnons, quantized magnetic excitations, across long distances in the low temperature easy-axis phase of the insulating AFM hematite (α-$Fe_2O_3$)[6,7]. In easy-axis AFMs, the excited magnons are circularly polarized, making them capable of transporting angular momentum, and thus information[8]. In the absence of a magnetic field, magnetic anisotropy, or some other symmetry-breaking mechanism, the two magnon modes are degenerate, carrying equal and opposite values of angular momentum and no net spin transport is observed. It has been shown that the application of a spin-bias at the interface of the AFM and a heavy metal can lead to an excess of magnons with one polarization, enabling net spin transport[6–8]. The efficiency of the magnon transport relies on a parallel alignment of *n* and the spin-bias and can be modified by the application of a magnetic field[6–8]. For AFM magnonic devices, one needs long-distance spin transport, conventionally thought to be facilitated by an antiferromagnet with an easy-axis anisotropy. However, the low temperatures required to stabilize the easy-axis phase of α-$Fe_2O_3$ makes it unsuited for AFM magnonic devices, which operate at room temperature. In the high temperature easy-plane phase, however, the low-energy magnons are, in general, predominantly linearly polarized. Very recently, the long-distance transport of magnon spin currents has been reported in the easy-plane phase of hematite, mediated by a superposition of linearly-polarized magnon modes that dephases[9–11] and is strongly suppressed as compared to the easy-axis phase[10]. To improve the efficiency of AFM magnon transport at room temperature alternative methods of altering the magnetic anisotropies are required. One possible way is to alter the magnetic anisotropies of α-$Fe_2O_3$ by the introduction of dilute dopants which offer increased opportunities for tuning the magnetic properties. The addition of dopants generally suppresses the easy-axis phase[12,13] and increase the conductivity of hematite[14,15], thus increasing the magnetic damping and reducing the magnon transport efficiency. As shown in Refs. 9-11, magnon transport is still possible in easy-plane anisotropy antiferromagnets but with a reduced efficiency compared to easy-axis AFMs and requires a magnon dispersion with a small separation between the active magnon modes in order to enable mode superposition with a large dephasing length. Whether the introduction of dopants, whilst maintaining the insulating nature of hematite, can aid in the transport of magnons, either by altering directly the magnetic structure or by changing the underlying magnon modes, and how the transport in such systems with complex anisotropies is affected by magnetic fields are open questions that are key to gauge the applicability of this material. If the introduction of dopants indeed allows for efficient magnon transport, this opens paths towards engineering of magnetic properties for AFM magnonic components without the need to maintain a transition to the easy-axis phase.

In this paper, we make use of high-quality thin films of hematite doped with Zn, to investigate the magnon transport at room temperature. The transport properties of these films are determined as a function of the applied magnetic field and propagation distance of the magnons allowing us to extract effective attenuation lengths. Even though the introduction of Zn significantly alters the magnetic anisotropy, we observe efficient long-distance magnon transport. We make use of magnetic imaging to spatially resolve the impact of dilute doping on the equilibrium orientation of the magnetic order in thin film hematite and find that the magnetic-anisotropy symmetry is neither easy-axis nor easy-plane. We find a surprising dependence of the transport on an applied field, which is explained by models based on coherent frequency coupling of magnons. These observations open new avenues for the tuning of magnon transport in thin film hematite for applications.

150-nm films (0001)-oriented hematite doped with 1% Zn (α-$Fe_{1.99}Zn_{0.01}O_3$:(Zn)α-$Fe_2O_3$) were grown by pulsed laser deposition (PLD) on (0001)-oriented $Al_2O_3$ substrates from a stoichiometric target as detailed elsewhere[16–18]. Despite previous reports of additional $ZnFe_3O_4$ phases[19–21], we observe purely the α-$Fe_2O_3$ phase (Fig. 1a) with a diffraction peak angle of 2θ=39.227°, slightly smaller than the bulk value[22] confirming previous reports on homogenous doping of hematite grown by PLD[18]. The inset of Fig. 1a shows the rocking curve of the (0006) hematite peak, indicating the highly epitaxial nature of our films. Unlike reported previously for other growth methods[19–21], insulating behavior is observed an electrical resistance of more than 100 GΩ, possibly due to Zn being divalent and replacing $Fe^{3+}$ within the hematite structure, as well an absence of a $ZnFe_3O_4$ phase for such a low concentration of Zn[20].

The magnetic state is characterized by magnetometry using a superconducting quantum interference device (SQUID), shown in Fig. 1b for a magnetic field applied in-plane, perpendicular to [0001], and out-of-plane, parallel to [0001], where a linear subtraction has been made to account for the diamagnetic substrate background. Within the plane perpendicular to [0001], the antisymmetric, Dzyaloshinskii-Moriya interaction[3,22] leads to a canting of *n* that can be detected in magnetometry measurements as a hysteresis for a magnetic field within this plane, see Fig. 1b. An out-of-plane field elicits a response typically of a hard-axis loop but with a different saturation magnetization. Reports of undoped hematite films have shown that parasitic layers of the ferrimagnetic γ-$Fe_2O_3$ can nucleate at the interface between the substrate and the α-Fe$_2$O$_3$[23] which may explain this difference that is not present in bulk hematite[3]. The x-ray absorption spectra (XAS) of the Fe edge is shown in Fig. 1c where clear linear dichroism (XMLD) is observed between linear vertically (V Pol.) and linear horizontally polarized (H Pol.) x-rays. The absence of x-ray magnetic circular dichroism (XMCD) confirms that our films are indeed antiferromagnetic at the surface, even with a possible ferrimagnetic seed layer. The canted moment is not visible in XMCD due to being below the available resolution. The symmetry of the XMLD and the subsequent measurements of the domain structure (see Fig. 3 and discussion) indicate that **n** lies predominantly in-plane[24].

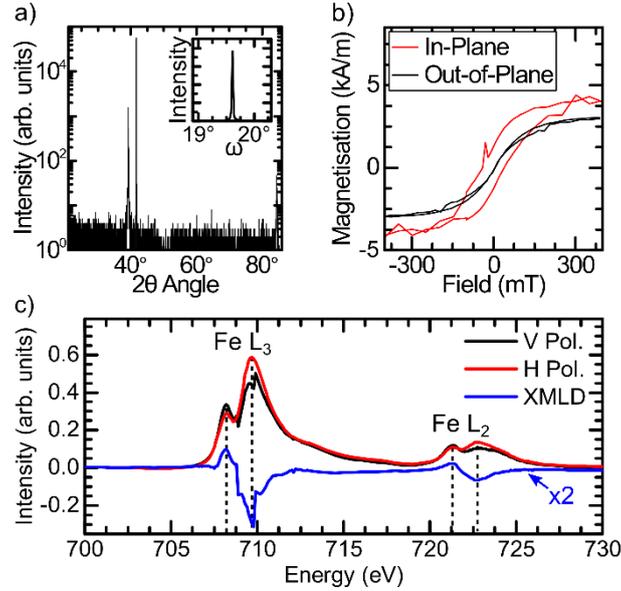

Figure 1. a) X-ray diffraction of (0001)-orientated, (Zn)α-Fe$_2$O$_3$ showing no signs of secondary phases. Inset: rocking curve of film (0006) peak showing the relaxed nature of the films. b) Magnetometry measurements via SQUID at 300 K for a magnetic field parallel (black) and perpendicular (red) to the crystallographic c-axis. c) X-ray absorption spectra for vertically (black) and horizontally (red) polarized x-rays at the Fe-L$_{2,3}$ edges. Clear linear dichroism (XMLD) is visible at both edges, multiplied by a factor of two to improve visibility.

Having established the high quality of the antiferromagnetic (Zn)α-Fe$_2$O$_3$ films, we now turn our attention to probing the magnon transport properties of these films to establish if the introduction of Zn significantly alters the magnetic damping or prohibits magnon transport completely. We fabricate 7-nm-thick Pt non-local devices by electron beam lithography (EBL), with electrical contacts made by a second EBL step and the deposition of Ti (4nm)/Au (32nm) (Fig. 2a). A charge current $j_c$ applied to one Pt wire leads to a spin accumulation $\mu_s$ at the Pt/(Zn)α-Fe$_2$O$_3$ interface due to the spin Hall effect (SHE)[25]. This spin accumulation creates a spin-bias across the interface, exciting AFM magnons in the (Zn)α-Fe$_2$O$_3$ due to exchange coupling across the interface[6–8,26]. Their polarization is determined by the orientation of $\mu_s$, which can be changed by reversing the current direction[6,7]. These excited magnons then diffuse away from the injector. They are absorbed by a second Pt wire placed at some distance $d$ and generate a spin current in the Pt that, by the inverse-SHE, creates a (negative) voltage $V$ that can be measured and converted to a non-local resistance $R_{el} = (V(j_c^+) - V(j_c^-))/2j_c$[6,7]. Very recent reports demonstrate that in easy-plane AFMs, a superposition of linearly-polarized modes can transport angular momentum, which dephases[9–11]. Alongside $R_{el}$, we can thus investigate the variation of the local resistance $R_L$ (see Fig. 2a), that varies due to the spin Hall magnetoresistance (SMR), where $R_L$ depends on the relative alignment of $\mu_s$ and the magnetic order parameters, the Néel vector **n**[2,3] and the canted moment **m**[27].

We start by investigating $R_L$ at 300 K for a magnetic field **H** applied along the length of the Pt wire (Fig. 2a), shown in Fig. 2b for increasing and decreasing **H**. We observe a sharp increase in $R_L$ at low magnetic fields before a saturation of the signal at the same field observed for the canted moment in Fig. 1b, indicating that $R_L$ reflects the magnetic state of the film. However, the increase indicates that here, $R_L$ is dominated by the reorientation of $m$ parallel to the magnetic field. If it was sensitive to $n$, this would arise as a decrease for this geometry[28]. A sensitivity to $m$ indicates that $R_L$ should also exhibit a hysteresis, so we investigate in the inset of Fig. 2b magnetic fields around 0 T. As expected, we observe a hysteresis in $R_L$, confirming that $R_L$ is dominated by the orientation of $m$ and the interfacial spin-mixing conductance associated with $n$ is significantly smaller than that associated with $m$[28]. This is consistent with an interfacial disorder-induced reduction in coupling to $n$[29], while the non-local spin transport measurements discussed further below demonstrated that a large coupling with the thermal magnon bath is maintained[30,31].

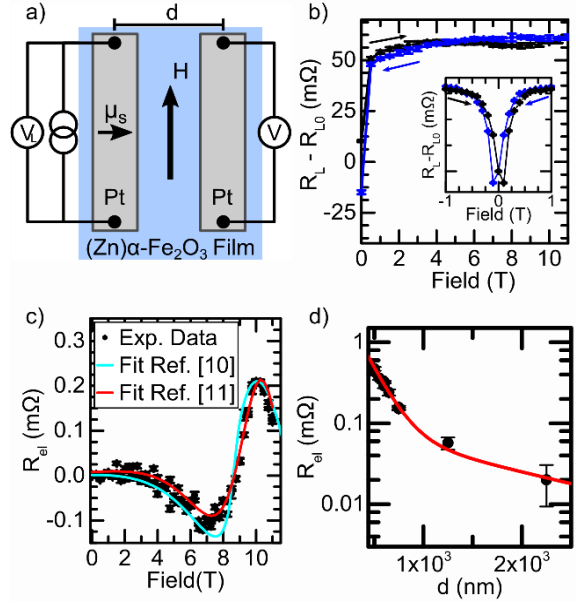

Figure 2. a) Schematic for non-local transport. Two Pt electrodes (grey) separated by a distance d (center-to-center) are patterned atop $(Zn)\alpha$-$Fe_2O_3$ films (blue). A charge current is applied to the left electrode, while a local and non-local voltage are measured. b) Local magnetoresistance of Pt/$(Zn)\alpha$-$Fe_2O_3$ as a function of magnetic field at 300 K for an increasing (black) and decreasing (blue) field. $R_{L0}$ is the average zero-field resistance. The inset shows fields around 0 T, where a hysteresis is visible. c) Non-local resistance as a function of field for a device with d=600 nm. The fits represent models centered around magnon mode-mixing [10,11] (details see main text). d) Decay with distance of the signal amplitude between the minimum and maximum values at $\mu_0 H$ = 6.8 T and 10.2 T, respectively. The fit accounts for the diffusive transport of magnons in the presence of an antiferromagnetic domain structure resulting in two exponential decays with $\lambda_1$=175±35 nm and $\lambda_2$=1.5±0.3 μm[7]. The error bars in c) and d) represent the standard deviation of the data points.

Next, we investigate the magnon transport at 300 K given that for application of AFM magnonic devices, room temperature operation is a key requirement. Previous magnon transport in pure AFMs has been seen at low temperatures in easy-axis antiferromagnets[6,7], where the key condition for magnon transport has been established as $n \parallel \mu_s$. At 300 K, undoped hematite is in the easy-plane phase[32], a phase that can be stabilized further by the addition of dopants[12,32], as indicated by our magnetometry and dichroism measurements in Fig. 1. In this phase, the excited magnon modes should be linearly polarized[33], reducing the potential to transfer angular momentum by a magnonic spin current. As the magnetic field is increased, $n$ will rotate to lie perpendicular to $H$ at some critical magnetic field in order to minimize the Zeeman energy, known as a spin-flop. Given that $m^2+n^2=1$ and $m \cdot n=0$, we would anticipate that $n \perp H$ occurs by 300 mT when the canted moment aligns to the field direction (see Fig. 1b and Fig. 2b), and $n \parallel \mu_s$. However, as seen in Fig. 2c, the application of $H \parallel j_c$ initially leads to no observable magnon transport and there is no evidence of the spin reorientation seen in Fig. 2b. As we increase $H$ further, we observe a negative signal, before a reversal at $\mu_0 H$ = 6.8 T. The signal begins to increase, before reaching a peak at $\mu_0 H = 10.2$ T, demonstrating that long-distance magnon transport is possible, even in doped hematite. For $H$ applied in-plane, perpendicular to the Pt wires, no spin transport is observed from the electrical excitation of magnons below the injector. When $H$ is applied out-of-plane, $n$ and $\mu_s$ are not aligned and no magnon transport occurs. This symmetry for $R_{el}$ indicates that the flowing magnon current is independent of $m$ and depends on the direction of $n$ only. However, a parallel alignment of $n$ and $\mu_s$ has already occurred for fields of only a few 100 mT, opening the question of what drives the spin transport in these films at high magnetic fields.

Recently, theoretical suggestions of the transport of angular momentum have been put forward where the uniaxial model is not applicable, relying on the coherent oscillations of the linear polarized eigenmodes[10,11]. Both models rely on the beatings that result from the breaking of axial symmetry about the Néel vector coherently coupling magnons with opposing polarization[11,34] and precession of pseudospin with frequency $\Omega$. However, the model of Ref. 11 considers the time-dependent beatings and bears a constant contribution from anisotropy and a linear-in-applied-field term stemming from antisymmetric exchange (DMI)[11]. In contrast, the model of Ref. 10 relies on the spatial beatings and associates the frequency $\Omega$ to the out-of-plane magnetic anisotropy and a term that is square in the applied field stemming from Zeeman energy. While a reduced magnetic symmetry due to doping renders our AFM films different from those investigated in Refs. 10 and 11, the two models can be applied to any anisotropy of the magnetic system. We show in Fig. 2c fits to $R_{el}$ making use of both models where we find good agreement despite the different regions of the Brillouin zone considered by each model (low k for the model of Ref. 10 and high k for the model of Ref. 11). This indicates that the full description of the spin transport should be based on a generalized model incorporating both models of Refs. 10 and 11. Further studies are required in order to disentangle the dominating transport mechanisms through varying the measurement geometry and environmental temperature. However, this is outside of the scope of the present study which shows that the observations here can be reproduced by both models.

Making use of a single distance, we obtain spin-diffusion lengths of 147±25 nm and 137±25 nm from the models of Ref. 10 and Ref. 11, respectively, in good agreement with the dominating spin relaxation length extracted from injector-detector distance dependent measurements (Fig. 2d).

To understand the transport mechanisms, we need to study the dependence of the non-local signal with wire separation, shown in Fig. 2d, by varying the separation between the wires. $R_L$ remains the same for all devices showing consistent interface properties. At each distance, the critical field where the transport is maximal is the same within 1% and no hysteretic behavior in $R_{el}$ is observed. Fig. 2d shows the distance dependence of $R_{el}$ at 300 K, where the value is taken as the difference between the minimum and the maximum of the signal in Fig. 2c. The signal drops exponentially with distance, indicating a dominating diffusive transport[6–8]. The exponential decay with distance of magnon transport can be described through a 1D-spin-diffusion equation taking the AFM domain structure into account[7]. The data in Fig. 2d can be fitted, as shown with the red line, with the two diffusion lengths $\lambda_1$=175±35 nm for the dominant scattering process and $\lambda_2$=1.5±0.3 µm for the higher frequency magnons that are not strongly affected by the domain structure[7]. The value of $\lambda_1$ gives an indication of the domain size[7], which we will investigate further later (Fig. 3), where we find an average domain size of 193±43nm. These length scales are also comparable to the spin-diffusion length found earlier, indicating that they are likely related. The dominant attenuation process, whether magnon dephasing or domain wall scattering, is not directly evident from this result.

As discussed earlier, we can explain our electrical measurements through a coherent frequency coupling of magnon modes. This relies on the assumption that the rotational symmetry is broken, requiring a magnetic state without a pure uniaxial anisotropy. To correlate the magnetic structure with the electrical measurements in Fig. 2, we perform XMLD-PEEM imaging at the Fe-$L_2$ edge at 300 K on identical films (Fig. 3a-3d). The samples are capped by 2-nm Pt to prevent charging. We observe large regions of contrast showing antiferromagnetic domains, where the contrast indicates the orientation of $n$. We do not observe any x-ray magnetic circular dichroism (XMCD), indicating that the canted magnetic moments are not visible within the experimental resolution. The average size of the domains visible is 193±43 nm, in good agreement with the estimated domain size from the non-local measurements (Fig. 2d) and also with the dephasing length based on the two theoretical models (Fig. 2c). Whilst the domain size may change with applied field, the absence of a training effect in the electrical measurements through repeated field-cycling can exclude dominating effects of this type[7]. This confirms that *n* in these domains facilitates magnon, and thus spin, transport. We highlight in Fig. 3, domains that have similar behavior as we rotate the incident x-ray angle φ or change the beam polarization. The presence of a uniaxial anisotropy is ruled out by the large domains with varying contrast given that easy-axis AFMs have 180° domains and thus no dichroic contrast. We also find that *n* has both an in-plane and out-of-plane component, differing between the highlighted domains. Consequently, *n* is not confined to the basal plane, and the magnetic anisotropy is more complex than the purely easy-plane anisotropy expected for undoped hematite[35]. Thus, we demonstrate that, by doping, one can control the magnetic properties whilst maintaining a good spin-transport medium.

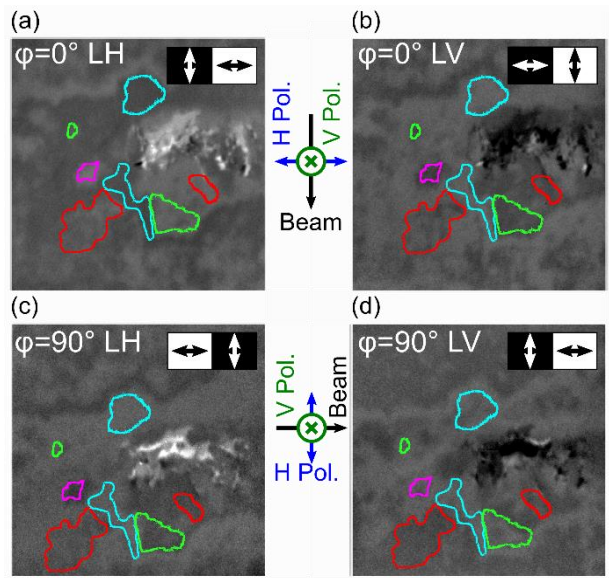

Figure 3. XMLD-PEEM images for (a) linear-horizontally and (b) linear-vertically polarized x-rays when the incident angle φ=0°. The object in the center is a non-magnetic particle used for alignment. XMLD-PEEM images for (c) linear-horizontally and (d) linear-vertically polarized x-rays for φ=90°. The highlighted regions indicate magnetic domains, where different colors indicate different behavior as a function of φ and beam polarization ruling out pure easy-axis and easy-plane anisotropy phases.

In conclusion, we have demonstrated long-distance magnon transport at room temperature in an antiferromagnet by doping thin films of α-$Fe_2O_3$ with Zn. Despite the introduction of Zn significantly altering the magnetic anisotropy and thus the equilibrium orientation of the Néel vector, the magnetic damping remains low enough to facilitate long-distance transport of magnons at room temperature without the need for a pure easy-axis or easy-plane phase. The low magnetic symmetry of the anisotropy caused by the additional dopants allows for magnon transport though mixing of differently-polarized magnon modes explained by models considering different regions of the Brillouin zone. The fact that two independent theoretical models reproduce our results indicates that the magnon transport is not confined to a single limit. The introduction of dilute dopants offers an attractive alternative for tuning the magnon-transport properties of thin film antiferromagnetic materials for room temperature applications.

**Data Availability**

The data that support the findings of this study are available from the corresponding author upon reasonable request.

**Acknowledgements**

A.R., S.D. and M.K. acknowledge support from the Graduate School of Excellence Materials Science in Mainz (DFG/GSC 266). This work was supported by the Max Planck Graduate Center with the Johannes Gutenberg-Universität Mainz (MPGC). A. R., R. L. and M.K. acknowledge support from the DFG project number 423441604. R.L. acknowledges the European Union's Horizon 2020 research and innovation programme under the Marie Skłodowska-Curie grant agreement FAST number 752195. R.L. and M.K. acknowledge financial support from the Horizon 2020 Framework Programme of the European Commission under FET-Open grant agreement no. 863155 (s-Nebula) O.G. and J.S. acknowledge the Alexander von Humboldt Foundation, the ERC Synergy Grant SC2 (No. 610115). All authors from Mainz also acknowledge support from both MaHoJeRo (DAAD Spintronics network, project number 57334897 and 57524834), SPIN+X (DFG SFB TRR 173, projects A01, A03, B02, and B12), DFG (423441604) and KAUST (OSR-2019-CRG8-4048.2). Av.R. acknowledges support from the European Research Council under the European Union's Seventh Framework programme (FP/200702013) / ERC (Grant Agreement No. 617516). D.A.G. acknowledges support from The Center for Absorption in Science, Ministry of Immigrant Absorption, State of Israel. The work including the Mainz-Trondheim collaboration was additionally supported by the Research Council of Norway through its Centres of Excellence funding scheme, project number 262633 'QuSpin'. L.B. acknowledges the European Union's Horizon 2020 research and innovation programme under the Marie Skłodowska-Curie Grant Agreement ARTES number 793159. We acknowledge Diamond Light Source for time on beamline I06 under proposal MM23819-1